\documentclass[12pt]{article}
\usepackage{amssymb,amsmath,amscd}
\usepackage{graphicx}
\makeatletter

\@addtoreset{equation}{section}
\def\section{\@startsection {section}{1}{\z@}{-2.25ex plus -1ex minus
 -.2ex}{1.0ex plus .2ex}{\large\bf}}
\def\subsection{\@startsection{subsection}{2}{\z@}{-2.0ex plus%
 -1ex minus -.2ex}{0.5ex plus .2ex}{\bf}}

\def\Ad{\mathrm{Ad}}

\def\bpi{{\mbox{\boldmath $\pi$}}}

\newcommand{\cd}[0]{\!\cdot\!}

\def\rlbicross{{\triangleright\!\!\!\blacktriangleleft}}

\def\dcross{{\bowtie}}

\def\bx{{\mbox{\boldmath $x$}}}

\def\bj{{\mbox{\boldmath $j$}}}

\def\bk{{\mbox{\boldmath $k$}}}
\def\bp{{\mbox{\boldmath $p$}}}

\def\bn{{\mbox{\boldmath $n$}}}

\newcommand{\ZZ}{\mathbb{Z}}

\newcommand{\RR}{\mathbb{R}}
\newcommand{\CC}{\mathbb{C}}

\def\bea{\begin{eqnarray}}
\def\eea{\end{eqnarray}}
\def\bmz{\left(\begin{array}{2,2}}
\def\emz{\end{array}\right)}
\def\bmd{\left(\begin{array}{3,3}}
\def\emd{\end{array}\right)}

\def\bpm{\begin{pmatrix}}
\def\epm{\end{pmatrix}}

\begin{document}
\parskip 6pt
\parindent 0pt
\begin{flushright}
EMPG-07-21
\end{flushright}

\begin{center}
\baselineskip 24 pt 
{ \Large \bf Lessons from (2+1)-dimensional quantum gravity}

\baselineskip 16 pt

\vspace{.5cm}
B.~J.~Schroers\footnote{\tt bernd@ma.hw.ac.uk} \\
Department of Mathematics and Maxwell Institute for Mathematical Sciences \\
 Heriot-Watt University \\
Edinburgh EH14 4AS, United Kingdom

\vspace{0.5cm}

{October    2007}

\begin{abstract}
\noindent  Proposals that quantum gravity gives rise to non-commutative
spacetime geometry and deformations of  Poincar\'e symmetry are examined in the 
context of (2+1)-dimensional quantum gravity. The results are expressed in five 
lessons, which summarise how the gravitational constant, Planck's constant
and the cosmological constant enter the non-commutative and non-cocommutative
structures arising in (2+1)-dimensional quantum gravity. It is emphasised
that the much studied  bicrossproduct  $\kappa$-Poincar\'e algebra does 
not arise directly in (2+1)-dimensional quantum gravity
\end{abstract}

{\footnotesize Based on talk given at the conference \\
``From Quantum to Emergent Gravity: Theory and Phenomenology''\\
  June 11-15 2007, Trieste, Italy}
 \end{center}               

\section{Introduction and motivation}

A key motivation for the study of (2+1)-dimensional  quantum 
gravity is to shed light on general and conceptual issues
associated with quantising gravity \cite{Carlipbook}.  
The goal of this talk is to focus on two closely related issues,
namely the role of non-commutative geometry 
and the emergence of deformed versions of special relativity in quantum gravity,
and to  extract lessons  regarding these issues from (2+1)-dimensional quantum gravity. 

In the course of the talk I
need  to refer to some  of the technical tools which make classical and quantum gravity in 2+1 dimensions
tractable, such as the formulation as a Chern-Simons theory and techniques from the theory of  quantum groups. However, I shall 
try to express each of the lessons in simple, physical terms. There will be a total of five lessons,
and all of them involve in an essential way the physical constants which enter quantum gravity,
namely the speed of light $c$, the gravitational constant $G$, the cosmological constant $\Lambda_c$
and Planck's constant $\hbar$. We will mostly set the speed of light to one, 
but exhibit the 
other constants explicitly. A special feature of 2+1 dimensions is that the Planck mass can be
 expressed in terms of $G$ only - without involving $\hbar$. The reason for this is that the dimension
 of $G$ in 2+1 dimensions is that of an inverse mass. It follows that we can form two length parameters 
\bea
\ell_P=\hbar G \qquad \ell_C=\frac{ 1 }{\sqrt{|\Lambda_c|}}
\eea
and one dimensionless ratio  $\ell_P/\ell_c$.

The talk is based on previous work with Catherine Meusburger \cite{we1,we2,we3} and on 
current work with Shahn Majid \cite{MS}. I begin be reviewing basic properties of the 
Poincar\'e group and special relativity in 2+1 dimensions. Promoting global  Poincar\'e symmetry
to a local symmetry leads to the formulation of gravity in 2+1 dimensions as a Chern-Simons gauge
theory. The cosmological constant can be introduced in this picture as 
a deformation parameter which 
changes the gauge group of the Chern-Simons theory. Quantisation 
deforms  the gauge group to a Hopf algebra which is neither commutative nor cocommutative.
However, one  important lesson  one learns in  2+1 dimensions is that the so-called bicrossproduct 
$\kappa$-Poincar\'e
algebra, much discussed in the recent literature on deformed or doubly-special relativity
(see e.g. \cite{KGintro1} for a review),
is not isomorphic to any of the Hopf algebras arising directly in the quantisation of 2+1 gravity,
contrary to what is sometimes claimed. I will conclude the talk with a 
careful explanation and discussion of this statement.

\section{Special relativity in 2+1 dimensions}
\subsection{Minkowski space and it symmetries}
I denote vectors in  three-dimensional 
Minkowski space by $\bx$ with coordinates $x^a$, $a=0,1,2$.
The metric is  $\eta_{ab} =$diag$ (+,-,-)$, so that the totally antisymmetric
tensor $\epsilon_{abc}$ satisfies $\epsilon^{abc} = \epsilon_{abc}$. 
The identity component of the Lorentz group is $SO^+(2,1)$, which is isomorphic to
$SL(2,\RR)/\ZZ_2\simeq SU(1,1)/\ZZ_2$. I denote the Lie algebra of the Lorentz group by
${\mathfrak su}(1,1)$ in this talk; its generators are the rotation generator $J_0$ and the boost generators
$J_1$ and $J_2$ with commutators
\bea
[J_a,J_b]=\epsilon_{abc} J^c.
\eea

The isometry group of Minkowski space  is the  Poincar\'e group, which
plays a key role this talk.  I will 
work with the double cover of the identity component of the Poincar\'e group
\bea
P_3= SU(1,1)\ltimes \RR^3
\eea
with multiplication law
\bea
(v_1,\bx_1)(v_2,\bx_2)=(v_1v_2,\bx_1+\Ad(v_1)\bx_2), 
\eea
where the notation exploits the identification of $\RR^3$ with the Lie 
algebra ${\mathfrak su}(1,1)$. The Lie algebra ${\mathfrak p}_3$ of the Poincar\'e group $P_3$ is generated by the Lorentz generators $J_a$ and translation generators
$P_a$, with brackets 
\bea
[ J_a,J_b ]=\epsilon_{abc}J^c, \quad [J_a,P_b]=\epsilon_{abc}P^c, \quad
[ P_a,P_b ] =0.
\eea
This  algebra has the invariant, non-degenerate inner product
\bea
\label{inprod}
\langle J_a,P_b\rangle = \eta_{ab}
\eea
which will be crucial in what follows.

\subsection{Phase space of a free point particle=(Co)adjoint orbit}

An excellent way to think about the phase space of any dynamical system 
is  as the space of all solutions of the equations of motion. For a free 
relativistic particle the phase space is then the space of all timelike
straight lines in Minkowski space. A given line can be parametrised by giving its
direction $\hat \bp$ and one  point $\bx$  on  it. Since the points $\bx $
and $\bx + \tau \hat \bp$ lie on the same line for any $\tau \in \RR$,
it is convenient to use, instead of $\bx$, the vector
\bea
\label{kpx}
\bk= \bx\wedge \bp + s\hat \bp,
\eea
for arbitrary but fixed $s \in \RR$.  Clearly, $\bk$ is 
 invariant under $\bx \mapsto \bx + \tau \hat \bp$. 
A given line is then uniquely characterised by two vectors $\bp$
and $\bk$ with fixed values for $\bp^2=m^2$ and $\bp\cd \bk=ms$.  The
space of all such lines (for given $m,s$)  is four dimensional.

\begin{figure}[h]
\centering
\includegraphics[width=10truecm]{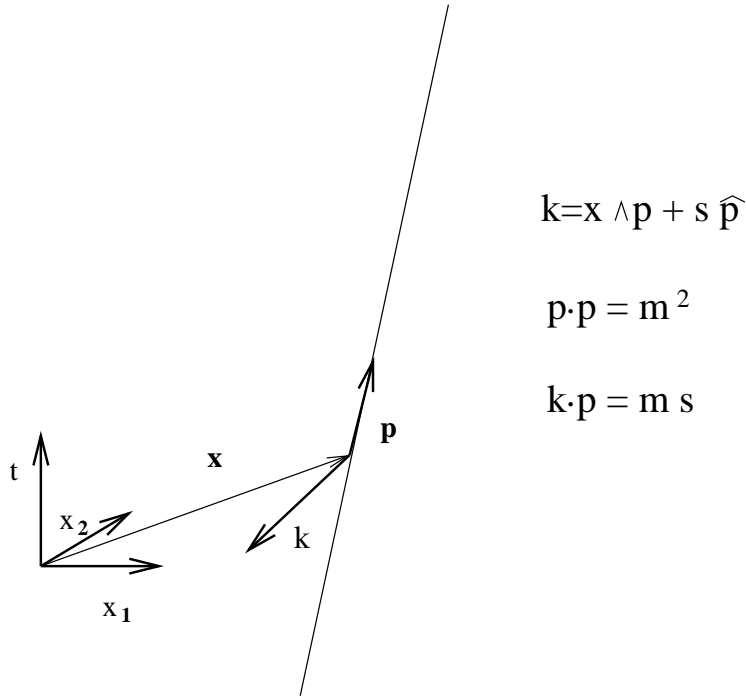}
\caption{Parametrising the world line of a particle}
\end{figure}

In order to derive the  symplectic structure on the 
phase space we require an action. In the case of the free relativistic
particle this action has a  geometrical interpretation   in terms of 
the coadjoint orbit method \cite{Kirillov}.  For convenience I use  the inner product
\eqref{inprod} to identify  the dual vector space ${\mathfrak p}^*_3$ 
with ${\mathfrak p}_3$ , and consider adjoint orbits instead of coadjoint orbits. In particular,
I identify 
\bea
 P^*_0\leftrightarrow J_0,\qquad 
J^*_0\leftrightarrow P_0
\eea
and  consider the  adjoint orbit of the Lie algebra  element $mJ_0 + sP_0$.
For $g=(v,\bx)\in P_3$ we define three-vectors $\bp$ and $\bk$ via
\bea
g(mJ_0 + sP_0)g^{-1} =p_aJ^a + k_aP^a,
\eea
which is equivalent to $p_aJ^a=\Ad(v)J_0$ and the relation \eqref{kpx}
between $\bx,\bp$ and $\bk$.
Then  the action  of a 
and spinning point particle  \cite{SousaGerbert} can be written as 
\bea
I_{\tiny \mbox{Point Particle}}= 
\int  d \tau \, p_a  \dot{x}^a +s\langle P_0,v^{-1} \dot{v} \rangle =
\int  d \tau \, \langle mJ_0 +s P_0, g^{-1}\dot{g}\rangle.
\eea
The resulting Poisson brackets are 
 the canonical Kirillov-Kostant-Souriau  brackets \cite{Kirillov} 
of the coordinate  functions $p_a$ and $k_a$:
\bea
\label{coadbrackets}
\{ k_a,k_b \}=-\epsilon_{abc}k^c, \quad \{k_a,p_b\}=-\epsilon_{abc}p^c, \quad
\{ p_a,p_b \} =0.
\eea

\section{2+1 gravity as a Poincar\'e gauge theory}
\subsection{The Chern-Simons formulation}
The starting point for the Chern-Simons formulation of 2+1 gravity is Cartan's trick of 
combining   the dreibein $e_a$  and spin connection $\omega = \omega_a J^a$
into the one-form
\bea
A=e_aP^a + \omega_aJ^a,
\eea
with values in ${\mathfrak p}_3$. 
As observed in   \cite{AT,Witten} 
 the Einstein-Hilbert action of 2+1 gravity can then be written as
a Chern-Simons action: 
\bea
\label{action}
I_{\mbox{\tiny Einstein-Hilbert}} = \frac{1}{8\pi G}\int_{M_3}  \, \langle A \wedge dA \rangle  + \frac{1}{3} \langle  [A, A] ,\wedge
 A \rangle.
\eea
Note that the definition of the action (but not of the connection) requires the inner product \eqref{inprod}.
The equation of motion following from \eqref{action} is the flatness condition for the curvature of the connection $A$:
\bea
F_A=0.
\eea
This is equivalent to requiring the spin connection to be flat and torsion free, and 
hence to the Einstein equations. 

\subsection{Introducing point particles}
We consider a  spacetime of topology
\bea
M_3= S_{gn}\times \RR, 
\eea
where $S_{gn} $ is a surface of genus $g$ with $n$ marked points, and introduce 
local coordinates $x=(x_1,x_2)$ on the surface $S_{gn}$ as well as a coordinate $\tau$ for $\RR$.
Each of the  
marked  points $S_{gn}$ is then decorated with a (co)adjoint orbit of the Poincar\'e group   which is  coupled 
to the gauge field via minimal coupling.  
Concentrating on one marked point, with coordinate $x^*$, the coupling is  
\bea
I_{\tiny \mbox{Point Particle}}= 
\int  d \tau \, \langle mJ_0 +s P_0, g^{-1}\left( \frac{d}{d\tau} +A_\tau (\tau,x^*)\right)g \rangle. 
\eea
The  equation of motion is now 
\bea
F_A=-g(\mu J_0 +\sigma P_0)g^{-1}\,dx_1\wedge dx_2\;\delta^2(x-x^*),
\eea
with $\mu = 8\pi m G$ and $\sigma = 8\pi s G $. This forces the holonomy around a given puncture
to lie in a fixed conjugacy class
\bea
 {\cal C}_{\mu \sigma}:=\{  g e^{-\mu J_0 -\sigma P_0} g^{-1}| g \in P_3\}.
\eea

\subsection{Holonomies and phase space}
In the Chern-Simons formulation, the phase space  of (2+1)-dimensional gravity can
be parameterised by holonomies around non-contractible loops on $S_{gn}$, see  Fig.~2 and \cite{we1}
 for further details. 
Defining the 
extended phase space via
\bea 
\label{extspace}
\tilde {\cal P} = P_3^{2g}\times {\cal C}_{\mu_n \sigma_n}\times \ldots {\cal C}_{\mu_1 \sigma_1},
\eea
the physical  phase space is obtained as a finite 
quotient:
\begin{align}
\label{pspace}
 {\cal P} & = \{(A_g,B_g,\ldots, A_1,B_1, M_n,\ldots M_1)\in \tilde {\cal P}| \nonumber  \\
   & \quad [A_g,B_g^{-1}]\ldots[ A_1,B_1^{-1}]M_n\ldots M_1=1\}/\mbox{conjugation}.
\end{align}
The space  $\cal P$  inherits a symplectic structure from the infinite-dimensional
affine space of connections $A$, of which it is an infinite-dimensional symplectic quotient.
The resulting symplectic structure on  the phase space $\cal P$ (called Atiyah-Bott structure) can be described
explicitly  in a framework introduced by Fock and Rosly \cite{FR}, and developed
in \cite{AMII} and \cite{AS}, see also   \cite{we1,we2} for  its 
application in (2+1)-dimensional gravity. The basic idea is to work on the extended
phase space $\tilde {\cal P} $,  and to define a symplectic structure on it in such a way
that the induced symplectic structure on the quotient \eqref{pspace} agrees with the Atiyah-Bott structure.
As emphasised particularly in \cite{AMII}, the Fock-Rosly symplectic structure on $\tilde {\cal P}$ 
is isomorphic, via a ``decoupling transformation'', to a direct product  symplectic structure consisting
of building blocks associated to the Poisson-Lie structure of the gauge group (for us $P_3$), 
namely a copy of the so-called Heisenberg double for every handle on $S_{gn}$, and  a symplectic
leaf of the dual or Semenov-Tian-Shansky structure for every particle. For details regarding 
this structures see \cite{AMI} and also \cite{K-S,CP} for further background. 

 \begin{figure}[h]
 \centering
 \includegraphics[width=11truecm]{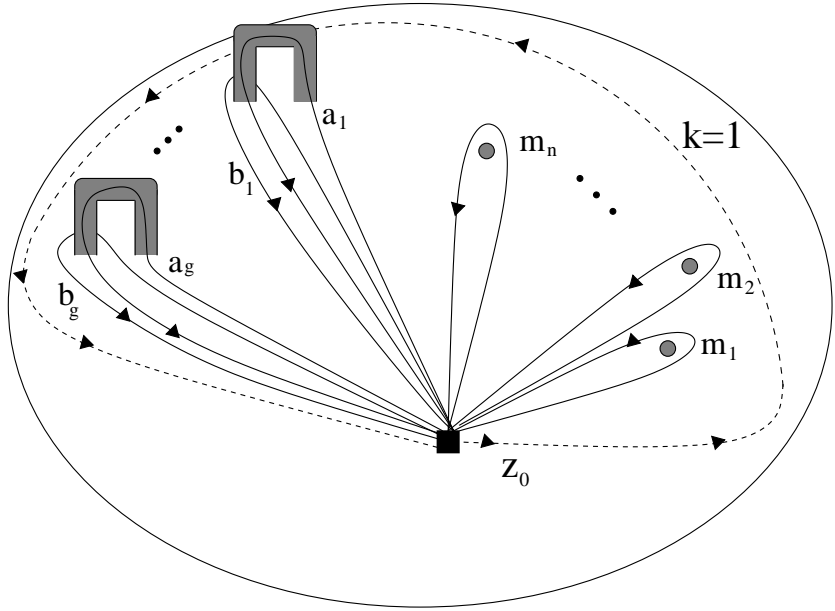}
\caption{The generators of the fundamental group of $S_{gn}$}
 \end{figure}

\subsection{$P_3$ as Poisson-Lie group}

A fundamental ingredient of the Fock-Rosly construction is 
an  $r$-matrix whose defining feature is 
that it satisfies the classical Yang-Baxter equation, 
and that its symmetric part agrees (after dualising) with the inner product \eqref{inprod} used in the definition of the Chern-Simons action.
It is easy to check that
the  r-matrix 
\bea
\label{rmat}
 r= P_a\otimes J^a\in {\mathfrak p}_3\otimes {\mathfrak p}_3.
\eea
satisfies these requirements. Given \eqref{rmat}, the group 
$P_3$  can be equipped with the Sklyanin bracket, thus turning it into a  
Poisson-Lie group. The bracket  takes the following 
form in terms of the parametrisation  $(v,\bx)\in P_3$: 
\bea
\label{Sklyanin}
\{x_a,x_b\}=G\epsilon_{abc}x^c, \quad \{x_a,f(v)\}=\{f(v),g(v)\}=0.
\eea

As mentioned above, it is not 
the Sklyanin bracket  itself  which enters the symplectic structure of the phase space but 
the associated Heisenberg double and dual Poisson structures.  We focus on the latter here,
and note that, as a group, 
the dual Poisson-Lie group  of $P_3$ is 
\bea
P_3^*=SU(1,1)\times \RR^3.
\eea
 To write down the Poisson structure of $P^*_3$ explicitly, 
 we write elements  as $(u,-\bj)$, with 
\bea
u=\exp(-8\pi G  p_aJ_a).
\eea
Then one finds  the following brackets of coordinate functions:
\bea
\label{dualbrackets}
\{j_a,j_b\}=-\epsilon_{abc}j^c, \quad \{j_a,p_b\}=-\epsilon_{abc}p^c, \quad
\{ p_a,p_b \} =0.
\eea
The Poisson manifold $P^*_3$ is a non-linear or deformed version of the  
linear Poisson manifold ${\mathfrak p}_3^*$.
The brackets \eqref{dualbrackets} 
are precisely the same brackets as those of  the coordinate functions
on ${\mathfrak p}_3^*$ \eqref{coadbrackets}. However, it is important to keep
in mind that for $P^*_3$,  the coordinates $p_a$ are functions on the non-linear
space $SU(1,1)$, whereas for ${\mathfrak p}_3^*$ they are 
functions on a  linear space.

\subsection{Conjugacy class as particle phase space}
One of the  results from  
the theory of Poisson-Lie groups which fits very beautifully into the current story is that the 
symplectic leaves of $P^*_3$ are  conjugacy classes in 
$P_3$.  We saw earlier that holonomies around a given puncture are forced to lie in 
 a fixed   conjugacy class   ${\cal C}_{\mu \sigma}$  of $P_3$, labelled by the 
mass and spin of the particle associated with the puncture. 
As shown in \cite{AMII}, the induced symplectic structure on those conjugacy classes
is precisely that of the dual Poisson-Lie group of the gauge group, in our case $P^*_3$.
The map between the conjugacy class in $P_3$ and the dual group $P^*_3$ is 
explicitly given by
\bea
(v,\bx)e^{-\mu J_0 -\sigma P_0}(v,\bx)^{-1}=(u,-\Ad(u)\bj)\mapsto (u,-\bj),
\eea
with the brackets between the coordinates $j^a$ and $p^a$ as in \eqref{dualbrackets}.
On the basis of those brackets we interpret $\bj$ as the ``angular momentum'' associated
to the particle,  and the element $u$ as a  ``group-valued momentum''.
We thus arrive at the following formulae for angular momentum and momentum in terms
of the Poincar\'e element $(v,\bx)$:
\bea
\label{pv}
 u&=&ve^{-\mu J_0} v^{-1} =e^{-8\pi G p_a J^a} \\
\label{jpx}
\bj &=& (1-\Ad(u^{-1}))\bx + s\hat p_a P_a 
= [\bx,p_aJ^a] + s\hat p_a P_a\; +\; {\cal O}(\bp^2).
\eea
The last line shows that the formula for $\bj$ can be viewed as a deformed
version of the relation \eqref{kpx} for a free relativistic particle.
Following this analogy we think of $x^a$ as position coordinates.
There is an important connection between the brackets \eqref{Sklyanin} of the position coordinates
and  the brackets \eqref{dualbrackets} of momentum and angular momentum: 
the conjugation action of $(v,\bx)$ on $(u,-\Ad(u)\bj)$ is a Poisson action
only if we take into account the non-trivial Poisson brackets of the  position coordinates $x^a$
given in \eqref{Sklyanin}. We thus arrive at 

\vspace{0.1cm}
\begin{center}

\begin{tabular}{|l|}
 \hline\\
\centerline{ \bf  LESSON 1: particle phase space in 2+1 gravity}\\
\\
$\bullet$ Momentum space has curvature radius  $\propto \frac 1 G $ \\
$\bullet$ Position coordinates do not Poisson commute $\propto G$  \\
$\bullet$ The angular momentum Poisson  algebra is unchanged - but the relation between  position, \\
\;\;\;momentum   and angular momentum is changed\\
\\
\hline
\end{tabular}
\end{center}

\section{Introducing the cosmological constant}
\subsection{Lie groups and Lie algebras}
In  2+1 gravity, solutions of the Einstein equations are locally
isometric to a model spacetime which is determined by the signature
of spacetime (Euclidean or Lorentzian) and the cosmological constant
\cite{Carlipbook}. The isometry groups of these  model spacetimes
are therefore
 local isometry groups in 2+1 gravity. In the formulation as a Chern-Simons
gauge theory \cite{AT,Witten}, the local isometry groups play the role of gauge groups.
We list the groups arising for different signatures and signs of the cosmological constant
in Table~1.

\begin{center}
\begin{tabular}{|c|c|c|}
\hline
  &    &   \\
Cosmological    & Euclidean signature & Lorentzian signature \\
constant & & \\
  &    &   \\
\hline
  &    &   \\
$\Lambda_c = 0$  & $E_3$ & $P_3$ \\
 &    &   \\
\hline
  &    &   \\
$\Lambda_c > 0$ & 
$ SO(4) \simeq  \frac{SU(2)\times SU(2)}{\ZZ_2}$&
$SO(3,1) \simeq SL(2,\CC)/\ZZ_2$ \\
  &    &   \\
\hline
  &    &   \\
$\Lambda_c < 0$ & $ SO(3,1) \simeq SL(2,\CC)/\ZZ_2 $&
$SO(2,2) \simeq \frac{SU(1,1)\times SU(1,1)}{ \ZZ_2} $\\
  &    &   \\
\hline
\end{tabular}
\vspace{0.4cm}

Table 1: Local isometry groups in 2+1 gravity

\end{center}

The Lie brackets of the associated Lie algebras can be written in unified fashion by
introducing
\bea
\Lambda = \left\{ 
\begin{array}{l l}
 \Lambda_c  & \mbox{for Euclidean signature} \\
   -\Lambda_c &\mbox{for Lorentzian signature}.
\end{array}
\right.
\eea 
They take the following form  in terms of generators $J_a$ and $P_a$
adapted to the 
Cartan decomposition:
\bea
[J_a,J_b]=\epsilon_{abc}J^c, \quad [J_a,P_b]=\epsilon_{abc}P^c 
\quad 
[P_a,P_b]=\Lambda \epsilon_{abc}J^c.
\eea
The  invariant pairing remains \eqref{inprod}
regardless of the value of $\Lambda$. Later we will also
need the  Iwasawa decomposition of the Lie algebras. As explained in   \cite{we3},
the generators
\bea
\tilde P_a=P_a+\epsilon_{abc}n^b J^c,  \qquad   \bn^2=-\Lambda, 
\eea
together with $J_a$ provide this decomposition. In particular one has
\bea
[\tilde P_a,\tilde P_b]=n_a\tilde P_b-n_b \tilde P_a.
\eea

It is explained in \cite{we3} how to write down Sklyanin, dual and Heisenberg
double brackets for the gauge groups listed in Table~1; as explained earlier,
this amounts to a complete description of the symplectic structure on the 
phase space in the Fock-Rosly framework. The cosmological constant introduces curvature into the model spacetimes of 
2+1 gravity; it is therefore not surprising that momenta, which generate
translations in the model spacetime,  no longer Poisson commute when the cosmological
constant is non-vanishing. 

\vspace{-0.5cm}
\begin{center}
\begin{tabular}{|l|}
\hline \\
\centerline{\bf  LESSON 2: the effect of the cosmological constant }\\
\\
$\bullet$ If $\Lambda \neq 0$  position space has curvature radius
 $\propto$ $\ell_c$ \\
 $\bullet$ $\Lambda \neq 0$ momenta do not Poisson commute 
$\propto \frac  1 \ell_c $ \\
$\bullet$ LESSON 1 still applies. \\
\\
\hline
\end{tabular}
\end{center}

\section{Quantisation}
\subsection{Quantisation of free  point-particle (coadjoint orbit)
brackets}
The quantisation of the Poisson algebra of momenta and angular momenta
of a free particle \eqref{coadbrackets}  leads to the 
 associative algebra generated by $J_0, J_1, J_2$ and $P_0,P_1,P_2$  
with relations
\bea
[J_a,J_b]=\hbar \epsilon_{abc} J^c, \qquad [J_a,P_b]=\hbar \epsilon_{abc}P^c,
\qquad [P_a,P_b]=0
\eea
The resulting algebra is the  universal enveloping algebra $U({\mathfrak p}_3)$ \cite{Dixmier}. 
Alternatively, one can think of the momenta as coordinate functions $p_a$ on momentum space $(\RR^*)^3$.
The ${\mathfrak su}(1,1)$  generators act on $(\RR^*)^3$ by infinitesimal rotations or boosts,
and hence on the polynomial  algebra Pol$((\RR^*)^3)$. One can view $U({\mathfrak p}_3)$ therefore also as
the semi-direct product of algebras
\bea
\label{freepaquant}
U(\mathfrak{su}(1,1))\ltimes \mbox{Pol}((\RR^*)^3).
\eea
The description of the momentum algebra as a function algebra offers certain
advantages which become manifest when one writes down the coalgebra
structure which turns \eqref{freepaquant} into a Hopf algebra.
The coalgebra structure encodes how momenta
and angular momenta of several particles are combined, see \cite{bamus} for details on this point of 
view.
For a free particle this is through simple addition i.e.
\bea
\Delta J_a= J_a \otimes 1 + 1 \otimes J_a
\quad \mbox{and} \quad 
\label{deltap}
\Delta (p_a)= p_a\otimes 1 + 1 \otimes p_a.
\eea
The last formula is a special case of the following  general construction.
Suppose $G$ is any Lie group, and  $\CC(G)$ is the abelian algebra of complex
valued functions on $G$, with pointwise multiplication\footnote{I do not discuss analytical aspects of this algebra in the current talk,
and therefore will not specify the class of functions further; however, we  do
require the functions to be differentiable}. Then we can define a
coproduct  via  
\bea
\label{comult}
\Delta:\CC(G)\rightarrow \CC(G\times G)
,\quad  \Delta \,f (g,h)=f(gh).
\eea 
For $G=(\RR^*)^3$  this leads to the rule  \eqref{deltap} for the coordinate functions $p_a$.

\subsection{The Lorentz  double}
It is explained in detail in \cite{we2}  how the  quantisation of the Poisson brackets  \eqref{dualbrackets}
of the momentum and angular momentum of a gravitating particle in 2+1 dimensions leads to the Hopf algebra
\bea
\label{lordouble}
D(U(su(1,1)):=U(\mathfrak{su}(1,1))\ltimes \CC(SU(1,1)).
\eea
This Hopf algebra is a particular example of a quantum double, and was 
called Lorentz double in \cite{bamus}. Following our discussion of the
phase of a particle in 2+1 gravity, it is not difficult to appreciate
how this algebra arises. The angular momentum algebra is unchanged compared
to the free relativistic particle, but the momentum coordinates are now
functions on the group manifold $SU(1,1)$ rather than the linear space $(\RR^*)^3$.
To go from \eqref{freepaquant} to \eqref{lordouble} we simply replace 
$\mbox{Pol}((\RR^*)^3)$ by the function algebra $\CC(SU(1,1))$.
Since the group $G=SU(1,1)$ is non-abelian it follows immediately that 
the momentum addition according to the general rule \eqref{comult} is not
cocommutative (i.e. depends on the order in the tensor product).
Applying the rule \eqref{comult} to group elements parametrised as in 
\eqref{pv}, and expanding in powers of $G$ one computes the leading 
order in non-cocommutativity. In Lesson 3 we combine this result with
the usual quantisation of the Poisson brackets \eqref{Sklyanin} for the 
position coordinates. Note that the lack of cocommutativity is 
independent of $\hbar$ and therefore really a classical effect;
it is merely the manifestation of the momentum space curvature 
in the language of the  Hopf algebra \eqref{lordouble}.
\vspace{-0.2cm}
\begin{center}
\begin{tabular}{|l|}
 \hline  \\
\centerline{\bf  LESSON 3: quantisation with vanishing cosmological constant} \\
\\
$\bullet $ $ [J_a,J_b] = \hbar \;\; \epsilon_{abc}J^c$: \;
Angular momentum coordinates do not commute $\propto \hbar$ \\ 
$\bullet $ $ \Delta(p_a)=1\otimes p_a + p_a\otimes 1 + G \;\;
\epsilon_{abc}p^b\otimes p^c +\ldots $:\;
Momenta do not cocommute $\propto G$ \\
$\bullet $ $[X_a,X_b] = l_P\;\; \epsilon_{abc}X^c$: \;
Position coordinates  do not commute $\propto l_P$\\
\\
\hline 
\end{tabular}
\end{center}

\subsection{Quantisation when $\Lambda \neq 0$}

The quantisation of 2+1 gravity has been been studied in the
so-called combinatorial or Hamiltonian framework
for the cases  $\Lambda=0$ (for both Euclidean and Lorentzian signature, see \cite{Schroers} and \cite{we2})
and $\Lambda <0$ \cite{BNR}. Table 2  lists  the quantum groups which (are believed to) play a role analogous to
that of the Lorentz double in the case $\Lambda =0$ (and Lorentzian signature).  For $\Lambda >0$ the 
quantisation along the lines described in this talk  has not been carried out in detail, so the  corresponding 
entries are conjectural. The  parameter $q$ in the table is
\bea
 q=e^{-\hbar G\sqrt{-\Lambda}},
\eea 
and combines all three physical constants
which enter 2+1 dimensional quantum gravity.
As usual,  $D(H)$ stands for the quantum double of a Hopf algebra $H$.

\vbox{
\begin{center}
\begin{tabular}{|c|c|c|}
\hline
  &    &   \\
Cosmological  const.  & Euclidean signature & Lorentzian signature \\
  &    &   \\
\hline
  &    &   \\
$\Lambda_c = 0$  & $D(U(su(2))$ & $D(U(su(1,1))$ \\
 &    &   \\
\hline
  &    &   \\
$\Lambda_c > 0$ & 
$ D(U_q(su(2)))$, $q$ root of unity &
$D(U_q(su(1,1)))$ $q\in  \RR$ \\
  &    &   \\
\hline
  &    &   \\
$\Lambda_c < 0$ & $D(U_q(su(2)))$, $q\in \RR$ & $D(U_q(su(1,1)))$,  $q\in U(1)$ \\
  &    &   \\
\hline
\end{tabular}

\vspace{0.3cm}
Table 2: Quantum groups arising  in 2+1 quantum  gravity
\end{center}
}
All of the quantum groups in Table~2 are non-commutative and non-cocommutative.
By studying the algebra and coalgebra structure of the quantum groups in Table~2, 
one can extract the parameters which control the failure of commutativity and 
cocommutativity to leading order. The results are summarised in the table  below;
it also includes a row for the position algebra, which I obtained by considering
the dual Hopf algebra of the momentum algebra. See \cite{BM} for a detailed discussion
of the postion algebra for the quantum double of $SU(2)$.

\vspace{0.3cm}
\begin{center}
\begin{tabular}{|c|c|c|}
\hline
  &    &   \\
{\bf LESSON 4}    & Commutator & Co-commutator \\
  &    &   \\
\hline
  &    &   \\
Angular momentum  & $\hbar$ &  $\frac  {G} {\ell_c} $ \\
 &    &   \\
\hline
  &    &   \\
Momentum &  $ \frac{\hbar}{\ell_c}  $ & $G$ \\
  &    &   \\
\hline
  &    &   \\
Position & $\hbar G $ & $\frac{1}{\ell_c}$ \\
  &    &   \\
\hline
\end{tabular}
\end{center}

\section{The $\kappa$-Poincar\'e algebra}

The so-called $\kappa$-Poincar\'e 
Hopf algebra was one of the first deformations of Poincar\'e symmetry 
proposed in the literature \cite{LNRT,MR}. At first sight, the $\kappa$-Poincar\'e algebra shares certain structural features with the Lorentz double:
 it has a   deformation parameter with the 
dimension of   mass,  it involves a curved momentum space, and it is isomorphic to the universal enveloping algebra
$U({\mathfrak  p}_3)$ as 
an algebra (though not as a Hopf algebra). 
As we have seen, momentum space
 in the Lorentz double is the group manifold $SU(1,1)$, which, as a Lorentzian manifold, is isomorphic
to anti-de Sitter space. In the standard version of the (2+1)-dimensional 
$\kappa$-Poincar\'e 
Hopf algebra, by contrast, momentum space is de Sitter space 
\bea
\mbox{dS}=\{(\bpi,\pi_3)\in \RR^4|-\pi_0^2+\pi_1^2+\pi_2^2 +\pi_3^2=\kappa^2\}.
\eea
The group $SU(1,1)$, and hence the Lie algebra ${\mathfrak su}(1,1)$, 
act  on de Sitter space. One can therefore define the
semidirect product of $ U({\mathfrak su}(1,1))$ with the algebra of complex-valued 
functions on $dS$:
\bea
U({\mathfrak su}(1,1))\ltimes  \CC(\mbox{dS}).
\eea
However, since  de Sitter space (unlike anti de Sitter space) is not a group manifold,
we cannot use our standard construction \eqref{comult} to define a coproduct.

In order to understand the construction of the coproduct we need to 
take a (short) detour  and review the bicrossproduct construction \cite{Majidbicross,Majid} of which 
the Hopf algebra structure of the $\kappa$-Poincar\'e algebra is a special case \cite{MR}.
The starting point of the construction is  
the  following factorisation of elements of the group $SL(2,\CC)$ (strictly speaking this only holds
for elements which obey  a certain condition, see \cite{we3} for details):
\bea
\label{factorise}
g\in SL(2,\CC)\Rightarrow g=u\cdot s= r\cdot v, \qquad
u,v \in SU(1,1) , \quad r,s \in AN.
\eea
where $AN\simeq \RR\ltimes \RR^2$ is group of matrices of form
\bea
r=\bpm e^{-\frac {p_0}{\kappa}} & \frac{p_1}{\kappa}+i
\frac{p_2}{\kappa} \\ 
0 & e^{\frac{p_0}{\kappa}}\epm.
\eea
Here $p_0,p_1,p_2$ are real parameter which we will eventually interpret as momentum
coordinates; the constant $\kappa$ has the dimension of mass, and is introduced 
at this point for purely dimensional reasons.
Next recall that (3+1)-dimensional Minkowski space can naturally be identified
with the vector space of Hermitian 2$\times$2 matrices, and 
that the action of  $g\in SL(2,\CC)$  on Hermitian matrices $h\mapsto ghg^\dagger$
implements (3+1)-dimensional Lorentz transformations. 
The de Sitter manifold can be realised  as a submanifold of the space of  Hermitian 
$2\times 2$ matrices via
\bea
\mbox{dS}=\{\pi_0+\pi_1\sigma_1+\pi_2\sigma_2 +\pi_3\sigma_3|
-\pi_0^2+\pi_1^2+\pi_2^2 +\pi_3^2=\kappa^2\},
\eea
where $\sigma_1,\sigma_2$ and $\sigma_3$ are the Pauli matrices.
Now note that dS is the orbit of $\kappa\sigma_3$ under the $SL(2,\CC)$ action,
and that the subgroup $SU(1,1)$ of $SL(2,\CC)$ is precisely the stabiliser
group of $\kappa\sigma_3$. Thus, provided the second factorisation in \eqref{factorise}
holds, we obtain a map
\bea
\label{ands}
AN\rightarrow \mbox{dS},\quad r\mapsto \kappa r  \sigma_3 r^\dagger.
\eea
In fact, as explained in \cite{we3}, the image of this map is only ``half''  of de Sitter space.
However, if we use the image of the map \eqref{ands} instead of all of de Sitter space as 
momentum space we obtain a curved momentum manifold which  has a group structure 
(that of AN). Moreover, since de Sitter space is acted on by (2+1)-dimensional
Lorentz transformations, we have an action of infinitesimal Lorentz transformations 
on the functions on ``half'' of de Sitter space.
Thus we can define 
\bea
 P_{\kappa}= U({\mathfrak su}(1,1))\rlbicross \CC(AN)    
\eea
which is a semi-direct product of algebras, and has a non-cocommutative momentum coproduct
\bea
\Delta(p_i)=p_i\otimes 1 + e^{-\frac{p_0}{\kappa}}\otimes p_i
\eea
which uses the group structure of $AN$. 
The symbol $\rlbicross $ indicates that there is 
a  twist in the  angular momentum comultiplication, 
but this will not concern us here.

\section{Relation with 2+1 gravity?}

We saw that the $\kappa$-Poincar\'e algebra is a bicrossproduct Hopf algebra; it has some 
 structural similarities with the Lorentz double, but is certainly not isomorphic to it. 
I will end this talk by sketching some observations about how these two Hopf algebras
can be related by a process called semidualisation \cite{Majid}.
Consider two Hopf algebras  which are each other's dual as Hopf algebra, e.g. 
\bea
\label{dual}
U_q({\mathfrak an})\leftrightarrow \CC_q(AN).
\eea
Semidualisation can be applied to Hopf algebras that factorise,
 and replaces one of the factors by its dual. For example, starting with
\bea
 U({\mathfrak sl}(2,\CC))\simeq  U({\mathfrak su}(1,1))\dcross 
U({\mathfrak an}), 
\eea
and using the classical ($q=1$) version of \eqref{dual}  the semidualisation map is  
\bea
 U({\mathfrak sl}(2,\CC)) \stackrel{S}{\rightarrow}U({\mathfrak su}(1,1))\rlbicross \CC(AN)=  P_\kappa 
\eea
Combining the semidualisation with 
the quantum duality principle \cite{Drinfeld,STS}
\bea
\CC_q(SU(1,1))\simeq U_q({\mathfrak an}),
\eea
which holds only when $q\neq 1$, we obtain the following diagram  \cite{MS}
\begin {center}
$
\begin{array}{ccccc}
U_q({\mathfrak su}(1,1))\dcross \CC_q(SU(1,1))  &\stackrel{q\neq 1}{\simeq}&
 U_q({\mathfrak sl}(2,\CC))&\stackrel{S}{\mapsto} & U_q({\mathfrak su}(1,1))\rlbicross \CC_q(AN)\\
&&&&\\
\downarrow q\rightarrow 1 &&&&\downarrow q\rightarrow 1\\
&&&&\\
D(U(su(1,1)))&&&& P_\kappa
\end{array}
$
\end{center}
\vspace{0.1cm}

Summarising the comparison between the bicrossproduct $\kappa$-Poincar\'e algebra and the quantum doubles arising in 2+1 quantum  gravity is the subject of the fifths and last lesson. Before coming to that summary,
I should comment on the suspiciously  vague word ``arising'' in
the previous sentence.
 
In this talk I have explained
the technical origin of  the quantum groups in Table~2 in the Fock-Rosly 
description of the phase space. However, the $r$-matrices used in the Fock-Rosly construction are really auxiliary objects, used to define a Poisson structure on the extended phase space \eqref{extspace}; the induced Poisson structure on the physical phase space 
only depends on the symmetric part of these $r$-matrices.
 Correspondingly, the quantum groups in Table~2 are auxiliary objects in the quantisation, and not uniquely associated 
to the quantum theory. Fortunately, there is
independent evidence that the quantum doubles of Table~2 play an
essential role in 2+1 quantum gravity, which does not make use of the Fock-Rosly
construction \cite{Noui}. Thus I think it is fair to say that 
2+1 quantum gravity does provide evidence for
the general idea that quantum gravity leads to a deformation of Poincar\'e
symmetry, with a deformation parameter of dimension mass; the Lorentz double
provides a specific realisation of this. 
I should stress that this is a deformation of Hopf algebras.
A meaningful discussion must take into account both
the algebra and the coalgebra structure. 
In the standard basis for the Lorentz double, for example,  the algebra
remains unchanged, and all the deformation takes place in  the coalgebra. 

By contrast, the role of 
the bicrossproduct $\kappa$-Poincar\'e Hopf algebra $P_\kappa$
in 2+1 quantum gravity remains, to my mind, unclear.
It is possible to obtain the $\kappa$-Poincar\'e algebra in 3+1 dimensions  by a  contraction procedure from $U_q(so(3,2))$
 in the limit $\Lambda \rightarrow 0$ \cite{LNRT}.    
This contraction procedure is sometimes interpreted 
as evidence for the emergence of $P_\kappa$ in a low energy limit of gravity in both 2+1 and 3+1 dimensions, see e.g.  \cite{ACSS}.
 However,  I am not aware of a careful version of this argument which  takes into account both the algebra and the co-algebra structure,
and  also keeps track of the $*$-structure (the analogue of a real structure for Hopf algebras, see e.g. \cite{Majid}). 
It is not sufficient to consider the algebra alone,
since the  bicrossproduct  $\kappa$-Poincar\'e  algebra, like the Lorentz double,  is  isomorphic to the  
Poincar\'e algebra  as an algebra (see e.g. \cite{KGN2}).  The $*$-structure matters because it distinguishes, for example,
$\mathfrak{su}(2)$ from $\mathfrak{su}(1,1)$, and therefore Euclidean from Lorentzian physics.
I indicated above another way of obtaining $P_\kappa$ by a sequence of mathematical
steps from one of the quantum doubles in Table~2; interpreting these steps physically and relating them
to the contraction procedure in \cite{LNRT} is the subject
of \cite{MS}. However, at this stage   the arguments
for a role of  $P_\kappa$  in 2+1 gravity seem far less convincing to me than those for the Lorentz double.

In relation to 3+1 dimensions, the situation in 2+1, as I see it, presents
a dilemma. The quantum groups which arise are all quantum 
doubles whose construction goes back to the essentially
 (2+1)-dimensional pairing \eqref{inprod}. Other constructions which
do generalise to higher dimensions, like the bicrossproduct construction,
by contrast,  do not arise naturally in 2+1 quantum gravity. 

\vspace{-0.4cm}

\begin{center}
\begin{tabular}{|l|}
 \hline 
\\
\centerline{ \bf LESSON 5: $\kappa$-Poincar\'e versus quantum doubles}\\
\\
$\bullet$ In 2+1 gravity momentum space is either Euclidean and positively curved 
 \\ \;\;\; (three-sphere) or
Lorentzian and negatively curved (anti-de Sitter).\\
\;\;\; The position algebra is $[X_a,X_b]=\ell_P \;\epsilon_{abc}X^c.$ \\
$\bullet$ In the  standard  bicrossproduct construction of $\kappa$-Poincar\'e,
momentum space \\
\;\;\; is Lorentzian  and positively curved (de Sitter).\\
\;\;\;  The position algebra is \;\;$[X_0,X_i]=\ell_PX_i.$\\
$\bullet$ Lorentz double and $\kappa$-Poincar\'e are 
{\em different} Hopf algebras  arising as $q\rightarrow 0$  
limits \\ \;\;\; of semidual Hopf-algebras .\\
\\
\hline
\end{tabular}
\end{center}

\end{document}